\documentclass{ws-rv9x6}
\usepackage{subfigure}     
\usepackage{ws-rv-van}     
\makeindex
\begin{document}

\chapter{Superfluid Local Density Approximation: A Density Functional Theory Approach to the Nuclear Pairing Problem }\label{SLDA}

\author[A. Bulgac]{Aurel Bulgac }

\address{Department of Physics\\
University of Washington, Seattle, WA 98195-1560, USA}

\begin{abstract}
I describe the foundation of a Density Functional Theory approach to include pairing correlations, which was applied to a variety of systems ranging from  dilute fermions, to neutron stars and finite nuclei. Ground state properties as well as properties of excited states and time-dependent phenomena can be achieved in this manner within a formalism based on microscopic input. 
\end{abstract}

\body

\section{Why use a Density Functional Theory approach? } 

The calculations of the ground and excited state properties of complex nuclei represent an ongoing challenge for several decades. Qualitatively we know a lot about these nuclear properties, however,  extracting these properties from a microscopic approach is still one of the most difficult problems in theoretical physics. The atomic nucleus presents a number of difficulties, not all of them shared by other quantum many-body systems. The existence of magic numbers is undoubtedly related to the existence of a well defined meanfield, in which nucleons move inside the nuclear medium. The experimental confirmation of the existence of superdeformed nuclei  and the semi-quantitative explanation of the equilibrium shape deformations based on semiclassical arguments represent another strong argument in favor of the existence of a well defined meanfield even in strongly deformed nuclei. Pairing correlations, which energetically represent a rather small decoration of the bulk nuclear binding energy, can equally well be described in terms of a generalized meanfield. At the same time the interaction among nucleons is strong and the description of the nuclear binding and structure in terms of the bare interaction among nucleons  is still a formidable challenge for the many-body theory, apart form brute force methods (quantum Monte Carlo (QMC) and no-core shell model) capable of handling so far only very light nuclei. In condensed matter theory a somewhat parallel approach was developed, in which the existence of well defined fermionic quasiparticle excitations played a fundamental role, the so called Landau Fermi liquid theory \cite{pines,gorkov}, subsequently extended to nuclei by Migdal and collaborators \cite{migdal}. Landau took it as a given that the ground state properties of strongly interacting systems cannot be calculated and have to be constructed in a phenomenological manner. In order to describe excited states one had to introduce additionally residual interactions and the emerging formalism was equivalent to the linear response theory \cite{pines}.  

In the 1970's a serious effort was mounted to calculate nuclear properties on an almost theoretically self-consistent approach, using effective interactions based on a G-matrix approach to nuclear matter \cite{bethe}. The approach was cumbersome, the theoretical errors were hard to evaluate, especially since a number of shortcuts (based on the current dominating prevailing theoretical ``intuition") were adopted. The use of the term ``effective" became so widespread that one can hardly find a universal meaning, often being used just as a misnomer for a semi-phenomenological approach of one kind or another.  Since the late 1970's and early 1980's we witness the powerful rise of the pure phenomenological approach to calculating ground and excited state nuclear properties using either Skyrme \cite{vautherin}, Gogny \cite{gogny} or relativistically inspired \cite{rmft} type of ``effective'' in-medium nuclear forces, with an immense proliferation of various parameterizations. There were a number of little nagging problems with all these models. The Skyrme interaction was formally interpreted as ``effective" in-medium nucleon-nucleon interaction, which can be used in a Hartree-Fock like calculation \cite{negele}. Certain contributions arising from this kind of interaction lead to undesirable effects especially in deformed nuclei and were unceremoniously dropped. Skyrme parametrization of the interaction was used only in the particle-hole channel (similarly in relativistic models), and a totally separate phenomenological approach was used in the particle-particle channel, in order to treat the pairing correlations. Theoretically one can bring hand-waving arguments that such an approach was meaningful, as a similar conclusion was reached in the more formal Landau Fermi liquid approach \cite{gorkov,migdal}, where one can clearly show that one obtains different contributions to the irreducible diagrams in particle-hole and particle-particle channels respectively, even though a number of sum rules leads to certain correlations among them. At the same time Gogny and his followers insist in using the same two-body ``effective" interaction in both the particle-hole and particle-particle channels,  even though it is hard to make a many-body theory argument in favor of such an approach. 

In the 21-st century the prevailing theoretical attitude changed.  It was realized and also became widely (though not unanimously) accepted that ``effective'' in-medium nuclear forces of one kind or another lack a clear theoretical underpinning (in particular these ``effective" in-medium forces are not observables and moreover cannot be defined uniquely), that their use was merely a means to obtain an Energy Density Functional (EDF) (which, however, can be phenomenologically parameterized directly), and that an entirely new theoretical concept makes more sense, namely that of an Density Functional Theory (DFT), formally introduced by Kohn, Hohenberg and  Sham in the 1960's for electron systems \cite{dreizler}.  DFT is in principle an exact approach, in which the role of the many-body wave function is replaced with the one-body density distribution, with the caveat that one needs to find an accurate energy density functional. The solution of the many-body Schr\"odinger equation is thus replaced with a significantly much simpler meanfield-like approach. DFT, particularly in its so called Local Density Approximations (LDA) of Kohn and Sham, proved to be widely successful in chemistry and condensed matter calculations of normal systems, where pairing correlation are absent. Fayans \cite{fayans} was perhaps the first to introduce DFT into nuclear structure calculations in a spirit very similar to condensed matter physics, namely by fitting an LDA functional to QMC results for infinite homogeneous matter, and adding a small number of phenomenological gradient correction terms. 

\section{DFT for a system with pairing correlations}

The DFT extension to superfluid systems has been performed in several ways so far, but by following a similar idea: one needs to add the anomalous density in order to describe the presence of a new order parameter. The first extension due to Gross and collaborators \cite{oliveira} added a dependence of the EDF on the nonlocal anomalous density matrix. In this way the great advantage of the LDA was nullified, as in this case one has to solve nonlocal or integro-differential equations. Especially in the case of nuclear systems, where pairing correlations are relatively weak and the size of the Cooper pair is larger than the radius of the nucleon-nucleon interaction, this kind of approach looks like an overkill. One expects that a large size and a weakly bound two-fermion Cooper pair could be accurately described  using a zero range interaction. One can easily show however that in such a case the emerging equations lead to an ultraviolet (UV) divergent anomalous density \cite{ab,by}. That was the main reason why Gross and collaborators  \cite{oliveira} resorted to a non-local extension of DFT to superfluid fermion systems. A natural UV-cutoff in electronic systems is provided by the Debye frequency. In nuclear systems the situation is quite different, as there are no phonon induced pairing correlations. 

A similar reasoning for dealing with UV-divergence was implicit as well in the Gogny parametrization \cite{gogny}, which uses a finite range ``effective" in-medium interaction, where the finite radius of the interaction provides the needed UV-cutoff in the particle-particle channel. In the particle-hole channel a finite range can be converted quite accurately (particularly in the case of  exchange terms) into a number contact terms with spatial derivatives \cite{negele,gebremariam}, using the density matrix expansion approach.  In Skyrme-like EDFs used in literature most of the time practitioners introduce an arbitrary cutoff parameter, which is often used as an additional phenomenological parameter.  The nuclear pairing gap is small and the size of the nuclear Cooper pair is rather large in comparison with the nucleon-nucleon interaction radius, and it is hard to make the case that a finite range of the interaction is needed and that is responsible for the stabilization of the calculations. The theoretical situation here is totally similar to the nucleon-nucleon interaction for energies below the $\pi$-threshold \cite{kaplan} where a rigorous treatment using contact interaction terms is easy and accurate as well to implement. 

The nature of the UV-divergence in the pairing channel was clarified a long time ago \cite{ab} and a simple regularization scheme was later introduced \cite{by}. In the case of two particles interacting with a finite range interaction the $s$-wave function for either bound states or scattering states behaves as $\propto 1/|{\bf r}_1-{\bf r}_2|$ outside the interaction range. In the case of scattering states the wave function at low energies is $\exp(i{\bf k}\cdot {\bf r})+ f(k)\exp(ik |{\bf r}_1-{\bf r}_2|)/ |{\bf r}_1-{\bf r}_2|$. One can show that the anomalous density matrix $\nu({\bf r}_1,{\bf r}_2)$  satisfies an equation almost identical to the Schr\"odinger equation for two interacting particles \cite{ab} and that  $\nu({\bf r}_1,{\bf r}_2) \propto \Delta(({\bf r})/|{\bf r}_1-{\bf r}_2| $ when $|{\bf r}_1-{\bf r}_2|\rightarrow 0$ (for a zero-range interaction, where ${\bf r}=({\bf r}_1+{\bf r}_2)/2$) and exactly this is the reason why the diagonal anomalous density diverges. This divergence and its amplitude has physical meaning and one cannot simply hide it under the rug by introducing a ill-defined cutoff. Fermi devised a very simple approach to deal with this kind of situation, without making recourse to either arbitrary UV-cutoffs or to fictitious finite range effects, by introducing a pseudo-potential \cite{fermi}. Fermi's pseudo-potential approach can be implemented in a straightforward manner in treating pairing correlations \cite{by}.  In particular, in order to describe the value of the $s$-wave pairing gap only one coupling constant is needed for both protons and neutrons, as expected from isospin invariance \cite{yb}.  

In a parallel approach a many-body perturbative approach to EDF, using renormalized bare NN (and NNN) interactions was suggested by Furnstahl and collaborators \cite{drut}. Only a few study cases have been considered so far and only in the absence of pairing correlations.
 
A quite successful DFT approach for fermion systems with pairing correlations has been developed and applied to a diverse sample of physical systems\cite{by,yb,slda,higgs,bfm,vortex,vortices,nvortex,qsw,ionel} and was dubbed the Superfluid Local Density Approximations (SLDA) as a direct generalization of the LDA approach of Kohn and Sham. In LDA/SLDA single-particle wave functions appear explicitly and often this type of approach is referred to as orbital based DFT.  In SLDA one constructs a local EDF in terms of various densities (spin degrees of freedom are not shown for the sake of simplicity)
\begin{equation}
\rho({\bf r}) = \sum_n |v_n({\bf r})|^2, \;\;\;
\tau({\bf r}) = \sum_n |\mbox{{\boldmath$\nabla$}} v_n({\bf r})|^2, \;\;\;
\nu({\bf r})  = \sum_n u_n({\bf r})v_n^*({\bf r}).
\label{eq:dens}
\end{equation}
If spin-orbit interaction is present an additional density should be added. The sums in the definition of these densities are all performed up to an UV-cutoff, in order to avoid the UV-divergence of the kinetic energy and anomalous densities. Even though an explicit UV-cutoff appears in the SLDA formulation, no dependence of the observables on this UV-cutoff exists once this cutoff is chosen appropriately. Both kinetic energy density $\tau({\bf r})$ and the anomalous density $\nu({\bf r})$ diverge in a similar fashion and their contribution to EDF is handled by introducing a well chosen counter term, and subsequently the entire formalism becomes divergence and counter term free \cite{by}. One can show that a unique combination 
\begin{equation}
\frac{\hbar^2}{2m_\mathit{eff}({\bf r})} \tau({\bf r}) - \Delta({\bf r})\nu^*({\bf r})
\label{eq:kin_an} 
\end{equation}
is divergence free, where $m_\mathit{eff}({\bf r})$ is the effective mass, $\Delta({\bf r}) = -g_\mathit{eff}({\bf r})\nu({\bf r})$ is the pairing gap and $g_\mathit{eff}({\bf r})$ is the renormalized position dependent coupling constant defining the strength of the pairing correlations \cite{by}. This implies that both kinetic energy and anomalous densities appear in EDF in this combination alone.   

Since DFT does not provide any constructive recipes for the EDF, any suggested approach needs validation. Unlike in the case of a phenomenological approach in this case agreement with experiment is not the proof that the championed approach is correct. One needs to explicitly show that the solution of the Schr{\"o}dinger equation for a many-body fermion system, in which pairing correlations are present and the corresponding DFT incarnation produce the same results for the ground state energy and ground state one-body density distribution. Fortunately, an extremely interesting system, which has mesmerized theorists across most physics subfields as well as experimentalists in cold atom physics, exists. This is the unitary Fermi gas (UFG), a system of fermions with spin-up and spin-down, interacting with a zero-range interaction and an infinite scattering length \cite{bfm,bcsbec}. The UFG has properties very similar to the properties of dilute neutron matter \cite{bfm,bcsbec,carlson,gezerlis} as envisioned by Bertsch in 1999 \cite{bertsch}.  In this case it is possible to calculate with relatively high controlled accuracy the energy and a number of properties of the ground state of a large series of such systems, with various particle numbers  with spin-up and down, both in the case of homogeneous systems and systems in external confining potentials. The results for the ground state properties of the homogeneous state - specifically the ground state energy, the pairing gap and the quasiparticle excitation spectrum (effective mass)  - are used to build the SLDA EDF. In the case of a UFG the only dimensional scale in the system is the inter-particle distance and this fact constrains the possible EDF structure. Simple dimensional arguments show that apart from three dimensionless constants $\alpha,\;\beta$ and $\gamma$ the (un-renormalized) EDF of an unpolarized ($N_\uparrow=N_\downarrow$) UFG has the simple structure:  
\begin{equation}
    \mathcal{E}_{SLDA}[n,\tau,\nu] =\frac{\hbar^2}{m}\left [ 
    \frac{\alpha}{2} \tau ({\bf r})+  \beta \frac{ 3(3\pi^2)^{2/3} }{10}n^{5/3}({\bf r})+
    \gamma \frac{|\nu({\bf r})|^2 }{n^{1/3}({\bf r})} \right ] ,
    \label{eq:edf_ufg}
\end{equation}
and an additional $V_{ext}({\bf r})n({\bf r})$ term, for an arbitrary external potential in which the system might or might not reside. As discussed above, the kinetic energy and the anomalous densities diverge, and a well defined renormalization procedure was devised \cite{ab,by,bfm}, which amounts to using these two densities in a unique combination in SLDA functional, see Eq. (\ref{eq:kin_an}). The infinite matter QMC calculations \cite{carlson} (where $V_{ext}({\bf r}) \equiv 0$ and $n({\bf r}) = const.$) provide enough information to determine the dimensionless constants $\alpha,\;\beta$ and $\gamma$. 

An independent series of QMC calculations of a large number of systems with various numbers of fermions $N_\uparrow$ and $N_\downarrow$ in an external harmonic trap \cite{blume} provide results for the ground state properties of inhomogeneous systems. At this point one can use the SLDA functional to predict the properties of these finite systems, see following table for a sample of results \cite{slda,bfm}:

\begin{table}[ht]
\begin{center}
\begin{tabular}[t]{r l l r}
  \multicolumn{4}{c}{A sample of ground state energies for unitary fermions in a harmonic trap}\\
  $(N_\uparrow, N_\downarrow)$ & $E_{QMC}$ & $E_{SLDA}$ & Error \\
  $( 2,  1)$ & $ 4.281\pm 0.004$ & $4.417$ & $3.2\%$\\
  $( 2,  2)$ & $ 5.051\pm 0.009$ & $5.405$ & $7\%$\\
  $( 3,  2)$ & $  7.61\pm  0.01$ & $7.602$ & $0.1\%$\\  
  $( 3,  3)$ & $ 8.639\pm  0.03$ & $8.939$ & $3.5\%$\\
  $( 4,  3)$ & $11.362\pm  0.02$ & $11.31$ & $0.49\%$\\  
  $( 4,  4)$ & $12.573\pm  0.03$ & $12.63$ & $0.48\%$\\
  $( 5,  5)$ & $16.806\pm  0.04$ & $16.19$ & $3.7\%$\\
  $( 6,  6)$ & $21.278\pm  0.05$ & $21.13$ & $0.69\%$\\
  $( 7,  6)$ & $24.787\pm  0.09$ & $24.04$ & $3\%$\\   
  $( 7,  7)$ & $25.923\pm  0.05$ & $25.31$ & $2.4\%$\\
  $( 8,  8)$ & $30.876\pm  0.06$ & $30.49$ & $1.2\%$\\
  $( 9,  9)$ & $35.971\pm  0.07$ & $34.87$ & $3.1\%$\\
  $(10, 10)$ & $41.302\pm  0.08$ & $40.54$ & $1.8\%$\\
  $(11, 10)$ & $45.474\pm  0.15$ & $43.98$ & $3.3\%$\\
  $(11, 11)$ & $46.889\pm  0.09$ & $45.00$ & $4\%$\\
\end{tabular}
\end{center}
\end{table} 
The degree of agreement between the QMC results for the finite inhomogeneous systems and the corresponding SLDA is a measure of the ability of the DFT to describe strongly interacting superfluid fermionic systems.   One should keep in mind that the accuracy of the QMC results is currently at the level of 5\% (the least accurate quantity being the pairing gap $\Delta_{UFG}(QMC)=0.504(24)\varepsilon_F$ and $\Delta_{UFG}(exp.)\approx 0.45(5)\varepsilon_F$ ), and that the QMC results for the infinite matter and finite systems were obtained by different groups, using somewhat different numerical approaches. Apart form this reduced sample of results presented in this table many other theoretical results  (thermodynamic properties, collective states, thermodynamic and quantum phase transitions) \cite{slda,bfm} and also extensive comparisons with results of many experiments are available in literature. The good quality of the agreement shown here, along with the theoretical arguments presented in favor of the SLDA functional  above lend strong support to the assertion that superfluid correlations in fermionic systems can be accurately described within the DFT approach.  The quality of the agreement between the QMC results for finite systems in harmonic traps and the corresponding SLDA results (in particular to even-odd staggering)  is even more surprising, as one might have expected that derivative corrections might exist, specifically a term $\hbar^2|\mbox{{\boldmath$\nabla$}} n({\bf r})|^2/m\,n({\bf r})$, proportional to an undefined dimensionless constant. One can make the argument that such a term can be expected if the interaction has a finite range, and also that the introduction of such a term would completely destroy the agreement between the SLDA and the corresponding QMC results for the finite UFG systems \cite{slda,bfm}. 

Very natural theoretical arguments allow us to extend the validity of the SLDA functional even further. By invoking local Galilean covariance one can show that in the un-renormalized SLDA functional   (\ref{eq:edf_ufg})  one has to add to perform the replacement 
\begin{equation}
\alpha \frac{\hbar^2}{2m}\tau({\bf r}) \rightarrow \frac{\hbar^2}{2m}\tau({\bf r}) + (\alpha-1)\frac{\hbar^2}{2m}\left [\tau({\bf r}) -\frac{{\bf j}^2({\bf r})}{n({\bf r})}\right ], 
\label{eq:galilean}
\end{equation}
where ${\bf j}(({\bf r})$ is the one-body current density \cite{slda,bfm}. A straightforward extension exists as well to the case of a polarized UFG, when $N_\uparrow \neq N_\downarrow$ \cite{slda,bfm}. By changing the ratio $N_\downarrow/N_\uparrow$ a UFG undergoes a number of quantum phase transitions, from a uniform superfluid to a Larkin-Ovchinnikov phase,  in which the order parameter oscillates in space, and further to a superfluid with relatively weak $p$-wave pairing and a normal state at very small temperatures \cite{bfm}.  The SLDA extension to include currents allows the description of excited states, in particular vortices, and of time-dependent phenomena. The existence of a DFT extension to time-dependent processes has been proven for quite some time \cite{runge}. The TDSLDA extension allowed so far the study of a large number of phenomena in a UFG system: the excitation of the Anderson-Higgs modes \cite{higgs}, in which the magnitude of the pairing field is excited with a large amplitude; the Anderson-Bogoliubov sound modes; the vortex structure \cite{vortex} and the dynamical generation of vortices and their non-trivial dynamics \cite{vortices}, in particular the first microscopic description of vortex crossing and reconnection in a fermionic superfluid leading to quantum turbulence predicted by Feynman in 1956;  the generation of quantum shock waves and domain walls in the collision of two UFG clouds \cite{qsw}. 

\section{ Nuclear DFT} 

The nuclear EDF is slightly more complicated, as one has to embed the dependence on proton and neutron densities \cite{yb,ionel}. The general principle however is very similar, the nuclear EDF has to satisfy all required symmetries: rotational and translational invariance, parity and isospin symmetry, gauge symmetry and Galilean invariance. We will not discuss here questions related to symmetry restoration, quantization of large amplitude motion, and the DFT stochastic extension, which were addressed recently elsewhere \cite{long}. An un-renormalized local nuclear EDF should have the following structure:
\begin{eqnarray}
\mathcal{E}_{SLDA}&=&
\frac{\hbar^2}{2m_p}\tau_p({\bf r}) + \frac{\hbar^2}{2m_n}\tau_n({\bf r}) +
\varepsilon_N[\rho_p({\bf r}),\tau_p({\bf r}), \rho_n({\bf r}), \tau_n({\bf r}), ...] \nonumber \\ 
&+& \varepsilon_S[\rho_p({\bf r})+\rho_n({\bf r})] (|\nu_p({\bf r})|^2+|\nu_n({\bf r})|^2)\nonumber \\
&+& \varepsilon_{S'}[\rho_p({\bf r})+\rho_n({\bf r})] (\rho_p({\bf r})-\rho_n({\bf r}))(|\nu_p({\bf r})|^2-|\nu_n({\bf r})|^2) \nonumber \\
&+& e^2\int d^3r' \frac{ \rho_{ch}({\bf r})\rho_{ch}({\bf r'})}{|{\bf r}-{\bf r'}|}
+\varepsilon_{xc}[\rho_p({\bf r}),\rho_n({\bf r})].
\label{eq:nedf}
\end{eqnarray}
where the isospin symmetry is broken by the small difference between the proton $m_p$ and neutron $m_n$ masses and the Coulomb interaction. The Coulomb interaction has a Hartree contribution in terms of the charge density $\rho_{ch}({\bf r})$ (which is different from $\rho_p({\bf r})$ mostly due to the finite proton size) and an exchange-correlation contribution $\varepsilon_{xc}[\rho_p({\bf r}),\rho_n({\bf r})]$, which is neither a simple Fock term nor its Slater approximation. Phenomenological studies of nuclear mass formulas show that a better fit can be typically obtained by neglecting the Coulomb exchange \cite{brown}, and a many-body analysis of  the Coulomb exchange in a nuclear medium show that its magnitude is significantly reduced \cite{vasya} and this effect is encoded in the term $\varepsilon_{xc}[\rho_p({\bf r}),\rho_n({\bf r})]$.  The term $\varepsilon_N[\rho_p({\bf r}),\tau_p({\bf r}), \rho_n({\bf r}), \tau_n({\bf r}), ...]$, where the ellipses stand for other densities such as spin densities and derivatives of $\rho_{p,n}({\bf r})$, should be a symmetric function of the proton and neutron densities in order to satisfy the isospin invariance. The first five terms should be chosen so as to describe correctly the QMC results for infinite neutron and symmetric nuclear matter. This is how Fayans has defined the first implementation of the Kohn-Sham DFT approach to nuclei \cite{fayans}. In the case of infinite matter various derivative terms give a vanishing contribution and their contribution to the nuclear EDF has to be determined from the properties of finite nuclei.  The term $ \varepsilon_S[\rho_p({\bf r})+\rho_n({\bf r})] (|\nu_p({\bf r})|^2+|\nu_n({\bf r})|^2)$ is the most important term describing the nuclear pairing correlations, which also satisfies isospin invariance and it has been used to describe odd-even effects in more than 200 nuclei with a simple constant volume pairing ($\varepsilon_S \equiv const.$) \cite{yb,ionel}. The same formalism was used  to demonstrate that a vortex in neutron matter develops a large density depletion at its core, a feature which controls the pinning mechanism \cite{nvortex}.  The additional term $\varepsilon_{S'}[\rho_p({\bf r})+\rho_n({\bf r})] (\rho_p({\bf r})-\rho_n({\bf r}))(|\nu_p({\bf r})|^2-|\nu_n({\bf r})|^2)$ is also isospin symmetric, but at this time it is not entirely clear whether such a term is present and whether its presence is required. In many phenomenological nuclear mass studies \cite{witek} various authors introduce independent coupling constants in the particle-particle channel for neutron and protons, which is a clear violation of the isospin symmetry. What is even more puzzling is the fact that these phenomenological studies claim that the strength of the coupling is stronger in the proton channel than in the neutron channel, which is hard to substantiate  microscopically. Another inconsistency of these phenomenological approaches is the introduction of different coupling strengths for even and odd systems \cite{witek}, which is not needed and is theoretically unjustified \cite{yb}. It is natural to expect that the pairing interaction is weaker in the proton channel, due to the repulsive character of the Coulomb interaction \cite{lesinski}. Another problem with approaches which try to be more microscopic in this respect \cite{lesinski,goriely} is the adoption of the pairing gaps calculated in the weak coupling BCS approximation, which are significantly higher than the actual values of the pairing gaps in infinite matter \cite{gezerlis}. It is known since 1961 \cite{heiselberg} that pairing gaps are reduced by a factor $(4e)^{1/3}\approx 2.2$, when one accounts for the contribution of the induced interactions, which is also confirmed by full QMC calculations of the UFG and of the neutron matter \cite{carlson,gezerlis}. Unfortunately there are no QMC calculations of the proton and neutron pairing gaps in symmetric nuclear matter, and thus the possible dependence of these gaps on the isospin composition of the nuclear matter is still unknown.  The divergences of the kinetic energy and anomalous densities in (\ref{eq:nedf}) should be dealt with as described above \cite{ab,by,slda,bfm}, see Eq. (\ref{eq:kin_an}). The Galilean invariance is retrieved by using the recipe described above in the case of kinetic energy density \cite{slda,bfm} by including proton and neutron current densities, see Eq. (\ref{eq:galilean}) and similar extensions for other densities \cite{brink}.  With the inclusion of currents the DFT can be used to describe excited states and time-dependent phenomena and collisions \cite{higgs,vortices,qsw,long,ionel}. The first application of the TDSLDA extension to a nuclear process was described recently\cite{ionel} by calculating for the first time the excitation of the giant dipole resonances in deformed open-shell heavy nuclei without any unjustified approximations and the agreement with experimental data is very good, without the need of any fitting parameters. Formally the solution of the TDSLDA equations appears as a time-dependent Hartree-Bogoliubov formalism in 3D and the complexity of these equations requires the use of leadership class computers, as it amounts to solving tens to hundreds of thousands of coupled nonlinear time-dependent 3D partial differential equations for tens to hundreds of thousands of time steps. With TDSLDA a microscopic treatment of low energy nuclear collisions, analogous to the collisions of superfluid atomic clouds performed recently \cite{qsw}, and induced nuclear fission in particular is within reach for the first time.

The author thanks his collaborators  M.M. Forbes, Y.-L. Luo, P. Magierski, K.J. Roche, V.R. Shaginyan, I. Stetcu, S. Yoon, and Y. Yu, and the support received from the US Department of Energy through the grants  DE-FG02-97ER41014 and DE-FC02-07ER41457.

\end{document}